\documentclass[11pt,twoside]{article}

\usepackage{asp2014}

\aspSuppressVolSlug
\resetcounters

\begin{document}

\title{Bright Star Subtraction Pipeline for LSST: Progress Review}

\author{Amir ~E. Bazkiaei,$^{1,2,3}$ Lee ~S. Kelvin,$^4$ , Sarah Brough,$^{3,5}$ Simon ~J. O'Toole,$^{1,2,3}$ Aaron Watkins$^6$ and Morgen ~A. Schmitz$^4$}
\affil{$^1$Australian Astronomical Optics, Macquarie University, North Ryde, NSW 2113, Australia; \email{amir.ebadati-bazkiaei@mq.edu.au}}
\affil{$^2$Astrophysics and Space Technologies Research Centre, Macquarie University, Macquarie Park, NSW 2109, Australia}
\affil{$^3$Australian Research Council Centre of Excellence for All Sky Astrophysics in 3 Dimensions (ASTRO 3D), Australia}
\affil{$^4$Department of Astrophysical Sciences, Princeton University, Princeton, NJ 08544, USA}
\affil{$^5$School of Physics, University of New South Wales, Sydney, NSW 2052, Australia}
\affil{$^6$Centre for Astrophysics Research, School of Physics, University of Hertfordshire, Hatfield, UK}

\paperauthor{Sample~Author1}{Author1Email@email.edu}{ORCID_Or_Blank}{Author1 Institution}{Author1 Department}{City}{State/Province}{Postal Code}{Country}
\paperauthor{Sample~Author2}{Author2Email@email.edu}{ORCID_Or_Blank}{Author2 Institution}{Author2 Department}{City}{State/Province}{Postal Code}{Country}
\paperauthor{Sample~Author3}{Author3Email@email.edu}{ORCID_Or_Blank}{Author3 Institution}{Author3 Department}{City}{State/Province}{Postal Code}{Country}

\begin{abstract}

\noindent We present the Bright Star Subtraction (BSS) pipeline for the Vera C. Rubin Observatory's Legacy Survey of Space and Time (LSST).
This pipeline generates an extended PSF model using observed stars and subtracts the model from the bright stars in LSST data.
When testing the pipeline on Hyper Suprime-Cam (HSC) data, we find that the shape of the extended PSF model depends on the location of the detector on the camera's focal plane.
The closer a detector is to the edge of the focal plane, the less the extended PSF model is circularly symmetric.
We introduce an algorithm that allows the user to consider the location dependency of the model.

\end{abstract}

\section{Introduction}

The unprecedented surface brightness depths that the Vera C. Rubin Observatory's LSST will reach when imaging the southern sky provides a spectacular opportunity for studying Low Surface Brightness (LSB) structures around galaxies \citep[e.g.,][] {2019ApJ...873..111I,2020arXiv200111067B}.
The presence of bright stars in such deep images can result in an over-estimation of the sky background if the bright stars are not robustly modeled to large enough radii and subtracted prior to background measurements.
Over-subtraction of the sky background destroys LSB structure in the type of deep imaging expected to be produced by LSST \citep{2020MNRAS.491.5317I, 2021ApJ...910...45M, 2023MNRAS.518.1195M}.
To prevent this, we are developing the Bright Star Subtraction (BSS) pipeline for LSST (Bazkiaei et al., in prep.).
The BSS pipeline will robustly model the stellar profile of bright stars out to several hundred arcseconds.
This extended Point Spread Function (PSF) model is subsequently used to subtract bright stars from deep images prior to sky background estimation.
A key benefit of this pipeline will be an increase in the effective area of LSST, avoiding the need to mask large areas around bright stars.
For example in the HSC Subaru Strategic Program (SSP) survey \citep{2018PASJ...70S...4A}, star masks remove $\approx$20$\%$ of survey area \citep{2022PASJ...74..247A}.

In this work, we use data from the HSC SSP survey to test our bright star subtraction pipeline.
HSC consists of 104 science detectors over a 1.5$^o$ diameter field of view \citep{2018PASJ...70S...1M}.
The Release Candidate 2 (RC2) dataset \citep{DMTN-091} is a subset of HSC data equating to $\approx$5 sq. deg. from $\approx$150 visits across 5 broad bands (\textit{g, r, i, z, y}).
The RC2 was constructed by the LSST Data Management (DM) team to provide a sufficiently large dataset with which to test ongoing algorithm changes to the LSST Science Pipelines.
As such, this makes it ideally suited for testing the Bright Star Subtraction pipeline for LSST.

\section{The Bright Star Subtraction Pipeline}

At present, the BSS pipeline utilizes three tasks from the LSST Science Pipelines\footnote{\url{https://pipelines.lsst.io/}} \citep{2018PASJ...70S...5B, 2019ASPC..523..521B}: \texttt{ProcessBrightStarsTask}, \texttt{MeasureExtendedPsfTask} and \texttt{SubtractBrightStarsTask}.
Each task is a highly configurable Python class.
The net result of this pipeline is the generation of a 2D extended PSF model (i.e. out to several hundred arcseconds) and the subtraction of the scaled model from bright stars in calibrated exposures.
The sections below discuss each task in more detail.

\subsection{\texttt{ProcessBrightStarsTask}}

This task takes calibrated exposures and the Gaia reference catalog \citep{2016A&A...595A...1G, 2023A&A...674A...1G, 2023A&A...674A..32B} as an input and produces normalized cutouts (`stamps') of stars.
Stars brighter than a user-defined magnitude limit are identified using the data butler \citep{2022SPIE12189E..11J}, and a cutout centered on the star with dimensions specified by the user is constructed.
The task then normalizes the flux level of the stamp using the value of the flux within a user-defined annulus centered on the star (`annular flux').
Optionally, a limit for the minimum fraction of valid pixels within an annulus can be specified, with only successful annuli preserved.
At the conclusion of this task, a single FITS file containing the normalized stamps and their mask planes is constructed.
Information on the stellar ID, magnitude and annular flux are persisted in the FITS header.

\subsection{\texttt{MeasureExtendedPsfTask}}

This task stacks normalized star stamps to generate an extended PSF model.
Various statistical methods for stacking are available, including median, mean, and sigma-clipped mean (default) with a user-defined outlier rejection sigma ($\sigma=4$ here).

We find that extended PSF models generated for individual HSC detectors depend strongly on the detector's location on the focal plane.
The further a detector is from the center, the greater the asymmetry of the extended PSF model (see Figure \ref{fig:hsc_focal_plane}).

\setcounter{footnote}{2}
\articlefigure[width=7 cm]{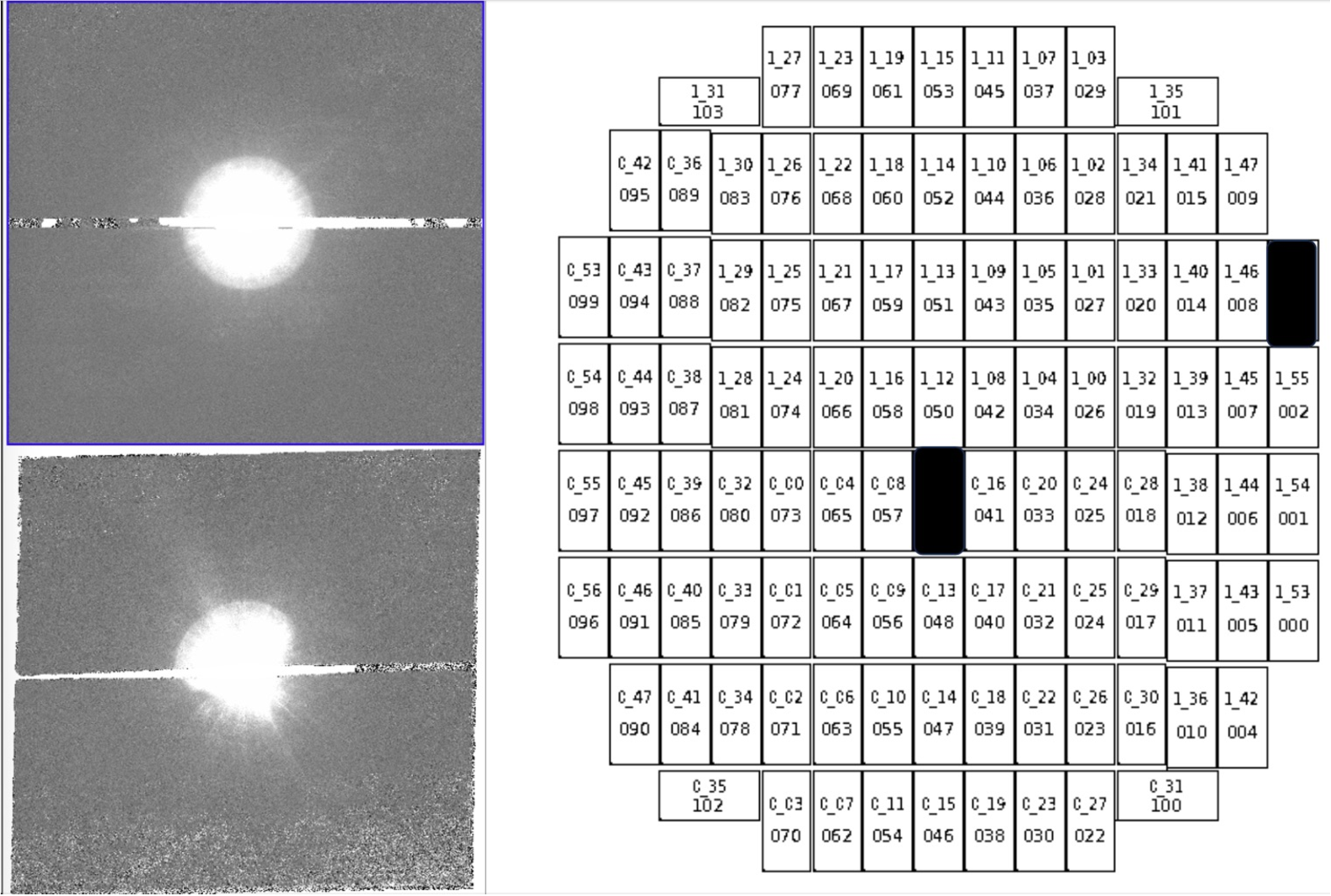}{fig:hsc_focal_plane}{
Top left: The extended PSF model for a central detector.
Bottom left: The extended PSF model for a peripheral detector.
The models are 1000 pixels on a side.
Right: The focal plane of HSC, consisting of 104 detectors (adapted from the Subaru Telescope website\addtocounter{footnote}{-1}\protect\footnotemark).
The numbers on each detector indicates the detector ID.
Filled rectangles on the focal plane show the location of the detectors which correspond to the extended PSF models in this figure.
Note how the detector near the center has a relatively circularly symmetric extended PSF model while the model corresponding to the detector near the edge shows a high level of asymmetry. 
}

\footnotetext{\url{https://www.subarutelescope.org/Observing/Instruments/HSC/hsc_ccd_anomaly.html}}

To address detector location dependency on the extended PSF model, the user can optionally define detector `regions' on the focal plane.
Each region may consist of one or more detectors, thereby allowing for an increase in signal-to-noise at the expense of a loss in some positional information.
Stamps from all detectors in a region are used to generate the per-region 2D extended PSF model.
These regional extended PSF model are then used to subtract bright stars from calibrated exposures (see below).

\subsection{\texttt{SubtractBrightStarsTask}}

This task scales generated a 2D extended PSF model to the brightness level of a star and subtracts it.
The Gaia reference catalog is used to identify all stars brighter than a user-defined threshold which are to be subtracted.
When all identified stars have been subtracted, a subtracted calibrated exposure is produced.
Each subtracted, calibrated exposure is stored in the data butler as a data product which may be retrieved as a FITS file on request.

Two extended PSF scaling types are provided: \texttt{annularFlux} and \texttt{leastSquare} (default) algorithms.
The \texttt{annularFlux} algorithm multiplies the extended PSF model by the annular flux of the star.
The \texttt{leastSquare} algorithm finds the least square scaling factor and multiplies the model by that factor.

Some stars may not ultimately be subtracted.
There are two primary reasons for this occurring.
First, these stars may have been flagged as a rejected star when the normalized stamps were originally produced using \texttt{ProcessBrightStarsTask}.
Second, the magnitude limit for stars to be subtracted may be fainter than the magnitude limit originally chosen to produce the normalized star stamps.

As rejected stars have no valid flux within the normalization annulus, this task iteratively tests a larger annulus to find a flux.
The first annulus at this step has an inner radius equal to the outer radius of the user-defined normalization annulus in \texttt{ProcessBrightStarsTask}.
The width of the annulus is conserved.
If no flux is found within this new annulus, the same procedure is repeated again to extend out to a larger annulus.
The search for a valid flux measurement continues following this recipe until the outer radius of the annulus reaches the edge of the stamp.
If no flux is found, the star will not be modeled nor subtracted.
For the second group of stars, the task carries out the same process but starts from the normalization annulus.
For each subtracted calibrated exposure, the task produces one output that contains stamps of unsubtracted stars.

\section{Works in progress and future works}

This paper describes the current status of the Bright Star Subtraction (BSS) pipeline.
Efforts are ongoing to improve the performance of the pipeline, add new functionality, and improve documentation.
Current topics of interest are exploring the use of multiple magnitude bins of stars to model different profile regions of the extended PSF.
We are also looking to extend our rejection methods (e.g. rejecting stars with bright neighbors), improve model scaling for stars near extended sources, develop methods for automatically and regularly measuring the fidelity of our results, and to add more unit testing capability.

\acknowledgements This material is based upon work supported in part by the National Science Foundation through Cooperative Agreement AST-1258333 and Cooperative Support Agreement AST-1202910 managed by the Association of Universities for Research in Astronomy (AURA), and the Department of Energy under Contract No. DE-AC02-76SF00515 with the SLAC National Accelerator Laboratory managed by Stanford University. Additional Rubin Observatory funding comes from private donations, grants to universities, and in-kind support from LSSTC Institutional Members.

This work has made use of data from the European Space Agency (ESA) mission
{\it Gaia} (\url{https://www.cosmos.esa.int/gaia}), processed by the {\it Gaia}
Data Processing and Analysis Consortium (DPAC,
\url{https://www.cosmos.esa.int/web/gaia/dpac/consortium}). Funding for the DPAC
has been provided by national institutions, in particular the institutions
participating in the {\it Gaia} Multilateral Agreement.

\bibliography{P924.bib}  % For BibTex

\end{document}